\documentclass[12pt,twocolumn]{article}

\usepackage[english]{babel}
\usepackage[utf8]{inputenc}
\usepackage[T1]{fontenc}
\usepackage{cite}
\usepackage{physics}
\usepackage{mathtools}
\usepackage{xcolor}
\usepackage{graphicx}% Include figure files
\usepackage{dcolumn}% Align table columns on decimal point
\usepackage{bm}% bold math

\usepackage{amsmath}
\usepackage{amssymb}
\usepackage{rotating}
\usepackage{lscape}
\usepackage{longtable}
\usepackage{authblk} 

\usepackage{hyperref}

\begin{document}

\title{Testing Bump in the Cosmological Power Spectrum Using Dwarf Galaxies} 
\author[1]{Maxim Zabelkin}%
\author[2]{Sergey Drozdov}
\author[2]{Oleg Skorikov}
\author[2]{Sergey Pilipenko}

\affil[1]{Faculty of Space Research, Lomonosov Moscow State University, Leninskie Gory 1, 119991 Moscow, Russia}
\affil[2]{P.N. Lebedev Physical Institute RAS, Profsoyuznaya St. 84/32, 117997 Moscow, Russia}

\twocolumn[
  \begin{@twocolumnfalse}
\maketitle

\begin{abstract}
We analyze the possibility of using observational data on nearby dwarf galaxies -- their luminosity functions and spatial distributions — to constrain deviations of the cosmological power spectrum from the standard one. Specifically, we consider a cosmological model with a "bump" in the power spectrum at a wavelength of 1.3~Mpc and a dimensionless amplitude $\mathcal{A}=2.0$. Such a spectrum is motivated by observations of an excess number of galaxies at high redshifts. The bump leads to a noticeable increase in the luminosity function in the range $-13 > M_B > -17$ at $z=0$. Comparison with observations constrains the bump amplitude to $\mathcal{A} < 0.25$ at a 3-sigma significance level for a wavelength of 1.3~Mpc. For wavelengths smaller than 0.8~Mpc, the bump manifests only in the luminosity function of dwarfs with $M_B > -14$.
\end{abstract}

\vspace{.5cm}

\end{@twocolumnfalse}
]

\section{Introduction}
%Currently, the $\Lambda$CDM cosmological model is supported by a wide range of observational data. However, there exist several observations that can be interpreted in favor of various extensions to this model. These include the missing satellites problem \cite{TBTF}, the cusp-core problem \cite{deBlok}, the Hubble tension \cite{Riess21}, and a possible excess of galaxies at high redshifts \cite{Naidu2022a,Naidu2022b,Lovell22}. One of the quantities available for modification in the standard cosmological model is the density perturbation power spectrum. It characterizes the amplitude of matter density deviations from the mean in the Universe as a function of spatial scale.

Although the $\Lambda CDM$ cosmological model is supported by a wide range of observational data, there are several mismatches in the literature, which point out to a necessity of a model expansion. These mismatches include the missing satellite problem \cite{TBTF}, the cusp-core problem \cite{deBlok}, the Hubble tension \cite{Riess21}, and a possible excess of galaxies at high redshifts \cite{Naidu2022a,Naidu2022b,Lovell22}. One of the expansion possibilities for $\Lambda CDM$ lays in changing initial density deviations power spectrum. This spectrum is defined as follows:

\begin{equation}
    P(k) = \langle \delta_\textbf{k} \delta_\textbf{k}^{*}\rangle,
    \label{eq:power_spectrum}
\end{equation}
where \textbf{k} -- wavelength vector, k -- its length, $\delta_\textbf{k}$ -- a Fourier image of a density perturbation field($\delta(\textbf{r}) = (\rho(\textbf{r}) - \langle\rho \rangle) / \langle \rho \rangle$), $*$ means complex conjugation, angle brackets mean ensemble average.

According to modern theoretical understanding, initial matter density inhomogeneities arose during inflation, and the simplest form of the initial spectrum is a power law with an index close to 1. Depending on the inflation model, the spectrum may contain various features \cite{Ivanov94,2023JCAP...04..011I}: "bumps" (peaks at certain scales), changes in the spectral index at some scales, etc. Therefore, measuring the initial power spectrum can test inflation models.

Modern constraints on the power spectrum are summarized in \cite{Chabanier19}. It is well constrained in the wavenumber range from $2\times10^{-4}$ Mpc$^{-1}$ to 2 Mpc$^{-1}$ (corresponding to wavelengths approximately 30 Gpc to 3 Mpc). On scales smaller than about 3 Mpc, constraints are weaker, making this range of interest for study. In \cite{Tkachev23}, power spectrum models with a bump were proposed that better describe high-redshift galaxy data from the James Webb Space Telescope than the $\Lambda$CDM model. The best-fit model features a bump at scale $k_0=4.7$ Mpc$^{-1}$ ($k_0=7$ $h$/Mpc, where $h=0.67$ is the dimensionless Hubble constant), corresponding to a wavelength of about 1.3 Mpc.

The presence of a bump leads to various potentially observable consequences \cite{Tkachev23,Pilipenko24,Tkachev24,2025PhRvD.112b3512W,Eroshenko25,Nadler25}. Simple Press-Schechter theory suggests that increasing perturbation amplitude initially increases the number of gravitationally bound objects (halos) in the mass range corresponding to the bump scale, producing a "bump" in the halo mass function. Connection between mass and $k_0$ in given by:

\begin{equation}
    M_0 \approx \langle \rho \rangle (2 \pi / k_0)^3,
    \label{eq:m-k0}
\end{equation}
where $\langle \rho \rangle$ -- average matter density in the Universe. These halos with the mass close to $M_0$ form earlier than in $\Lambda$CDM. Over time, the halo mass function approaches the standard one, smoothing and shifting the bump toward larger masses. If halos formed due to the bump can form stars, this leads to early galaxy formation, affecting galaxy luminosity functions, the rate of reionization, and the 21 cm line signal. Increased numbers of dwarf halos may also affect expected dark matter annihilation signals.

In this work, we examine the bump's effect on dwarf galaxy counts in the nearby Universe at $z=0$. Dwarf galaxies have masses up to $3\times10^{10}$ $M_\odot$, corresponding to the bump scale with $k_0 = 4.7$ $\text{Mpc}^{-1}$. Such mass objects should have formed sufficiently early, at $z \ge 8$, corresponding to a Universe age under 1 billion years. The number of galaxies at this epoch was estimated in \cite{Tkachev23}. Over subsequent billions of years, galaxies underwent complex evolution, including mergers and star formation, not described by simple linear theory. The dark matter halo mass function across multiple epochs was also estimated in \cite{Tkachev23}, showing that by today the bump's effect on halo mass function is relatively small — information about the bump is lost: a bump that increased halo counts by ~70 times at $z=10$ for $3 \times 10^{10} M_\odot$ halos only increases counts by a factor of 3 at $z=0$. However, in this estimate baryonic evolution was not considered. Measuring virial masses for halos with $3 \times 10^{10} M_\odot$ or less is complicated observationally \cite{Zasov17}, and baryonic processes can make galaxies nearly unobservable due to low luminosity or surface brightness (e.g., ultra-diffuse galaxies \cite{vanDokkum15}). Thus, a factor of 3 difference in halo counts may be small enough to require further study to determine observability.

Unlike dark matter, which interacts gravitationally only, baryons undergo complex processes currently not fully modeled. These include gas cooling, ionization changes, chemical reactions, magnetic field interactions, star formation, and interactions with stellar radiation. Semi-analytic models, sets of differential equations approximating gas and star evolution based on dark halo growth history, are one approach to include these processes. These equations are based either on simple physical considerations, or approximate the results of detailed simulations.

We use the GRUMPY model \cite{Kravtsov22}, which has relatively few equations and an open-source Python code, facilitating understanding and modification. It was designed for dwarf galaxies and excludes processes important for massive galaxies, such as central supermassive black hole effects. \cite{Kravtsov22} shows it reproduces observed dwarf galaxy correlations: mass-size, mass-metallicity, and mass-star formation rate relations. \cite{Dekker25} used GRUMPY to constrain the primordial power spectrum, but considered spectra with altered slopes at small scales rather than bumps.

In order to test cosmological model with modified power spectra we model luminosity functions of dwarf galaxies using cosmological numerical simulations for the dark component and enhancing them by GRUMPY. Observational data, e.g. from \cite{klypin15}, show the dwarf galaxy luminosity function in the Local Volume follows a Schechter function for absolute magnitudes $M_B < -14$, so only models without strong features in this luminosity range are acceptable.

As an additional observational test, we analyze the two-point 3D galaxy correlation function (see, e.g., \cite{Hawkins03}). It is directly related to the power spectrum by Fourier transform, but on $\sim1$ Mpc scales, large-scale structure is nonlinear, especially in baryonic galaxy distribution, so the power spectrum derived from the correlation function differs from the initial one. To study the correlation function in the bump model and its difference from the standard spectrum, we compute it for model galaxy samples from GRUMPY. To our knowledge, observed dwarf galaxy correlation functions in the local neighborhood have not yet been obtained, but increasing galaxy samples will enable this.

\section{Numerical Dark Matter Halo Models and the GRUMPY Model}

In this paper we use two cosmological numerical simulations, one with standard spectra and one with modified. Due to impossibility of simulating the infinite Universe, simulation took place in a cube with a size equal to 47 Mpc(= 32/h Mpc) with periodic boundary conditions. This cube was filled by $1024^3 \approx 10^9$ dark matter test particles, particle mass $3.9 \times 10^6 \textbf{M}_\odot$.

%We used two numerical models with a cube size of 47 Mpc ($=32/h$ Mpc) filled with $1024^3 \approx 10^9$ dark matter test particles, particle mass $3.9 \times 10^6 M_\odot$. 

One model used the $\Lambda$CDM power spectrum; the other used the \textit{gauss\_1} spectrum from \cite{Tkachev23} featuring a bump at $k_0=4.7$ Mpc$^{-1}$ (wavelength 1.3 Mpc), bump width $\sigma_k=0.1$, amplitude $A=20$. The power spectrum differs from $\Lambda$CDM as:
\begin{equation}
\frac{P_\mathrm{bump}(k)}{P_{\Lambda \mathrm{CDM}}(k)} = 1 + A \cdot \exp\left(-\frac{(\log(k)-\log(k_0))^2}{\sigma_k^2}\right).
\label{eq:bump}
\end{equation}

It is important to note that, despite the fact that (\ref{eq:bump}) has three parameters, for narrow bumps with $\sigma_k \ll 1$, halo mass distributions and evolution depend only on $\mathcal{A} = A \sigma_k$. This can be shown using the Press-Schechter formalism as done in \cite{Tkachev23,Eroshenko25}. Here, $\mathcal{A} = 2$, required to explain the excess number of galaxies at $z \sim 10$ observed by JWST better than $\Lambda$CDM.

Initial conditions were prepared with Ginnungagap code \cite{ggp}. Simulations were run with public Gadget-2 code \cite{gadget}. Halo catalogs were extracted using Rockstar \cite{Behroozi_2013}, selecting halos with mass $>3 \times 10^8 M_\odot$ at simulation end. Merger trees were built tracing the most massive progenitor at each output time. Only the halos traced to $z>0.5$ were considered, yielding ~600,000 growth histories for the bump model and 1.2 million for $\Lambda$CDM.

\begin{figure}
    \centering
    \includegraphics[width=0.49\textwidth]{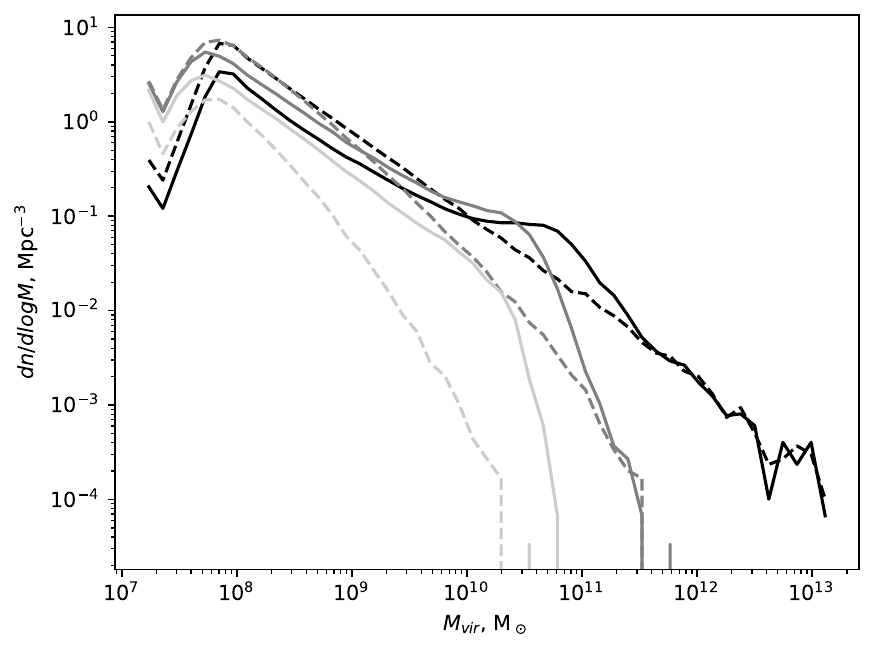}
    \caption{Dark matter halo mass functions at $z=0,6,11$ (dark to light lines). Solid lines: bump model; dashed lines: $\Lambda$CDM.}
    \label{fig:MF}
\end{figure}

Figure~\ref{fig:MF} shows dark matter halo mass functions. The resolution limit is $M > 10^8 M_\odot$ (halos with more than 30 particles). The bump's impact on the mass function (discussed in \cite{Tkachev23}) is evident: a significant increase at $z=11$ compared to the no-bump spectrum, and a slight increase at $z=0$ around $10^{11} M_\odot$ with a slight decrease below $10^{10} M_\odot$.

GRUMPY \cite{Kravtsov22} is a semi-analytic "regulator"-type model with a minimal set of differential equations describing atomic and molecular gas, stars, gas disk size, and metallicity evolution. It accounts for UV background effects (external and stellar), modifying gas accretion rates onto low-mass halos and star formation rates, as well as gas outflows. GRUMPY reproduces various dwarf galaxy relations seen in hydrodynamic cosmological models and observations.

The model outputs histories of stellar, molecular, and neutral gas masses. Luminosities are computed using star formation rates, assigning a Kroupa initial mass function \cite{Kroupa01} to new stars, calculating stellar mass distributions at observation time, and then using the FSPS stellar spectra library \cite{Conroy09,Conroy10} to obtain luminosities in various filters.

\section{Dwarf Galaxy Luminosity Function}
In the bump cosmological model (\ref{eq:bump}), the number of halos at $z \gtrsim 10$ increases in a certain mass range compared to the standard spectrum. By now, many halos have merged into larger objects. We check how this affects the observed galaxy number as a function of luminosity.

\begin{figure*}
    \centering
    \includegraphics[width=\linewidth]{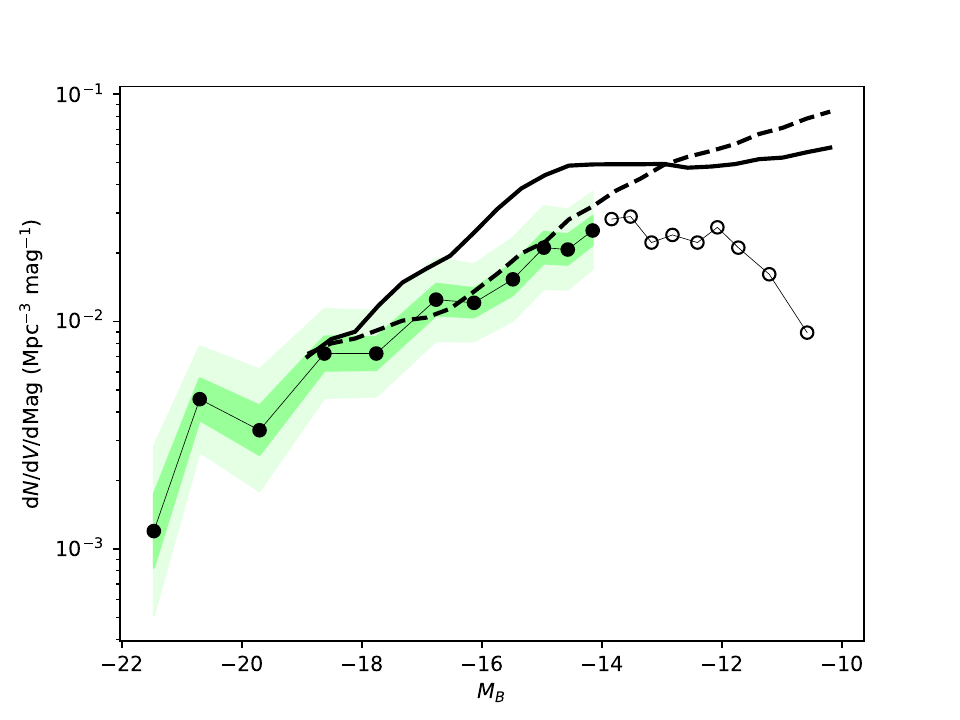}
    \caption{Dwarf galaxy luminosity function in the bump model (solid line) and standard spectrum model (dashed line). Circles show measurements from \cite{klypin15}. Filled circles correspond to $M_B < -14$, where the galaxy sample is complete. Shaded areas indicate Poisson errors at $1\sigma$ and $3\sigma$.}
    \label{fig:lum}
\end{figure*}

Figure~\ref{fig:lum} shows the luminosity functions in the B filter for bump and standard models. At $z=0$, the bump model yields fewer faint galaxies with $M_B > -13$ than the standard model. Around $M_B \approx -15$, the bump model shows a 3-fold excess compared to the standard, causing a visible kink in the luminosity function.

Observed dwarf galaxy luminosity function in the Local Volume is published in \cite{klypin15}. The sample is complete for $M_B < -14$ and fits a Schechter function well in this range. Figure~\ref{fig:lum} shows the standard model matches observations well, while the bump model exceeds observational error margins. The $\chi^2$ per degree of freedom for the standard model difference from observations is 1.1 (probability 0.32), while for the bump model it is 14 (probability $\sim 3 \times 10^{-41}$). Thus, the dwarf galaxy luminosity function disfavors a bump with parameters as above at 1.3 Mpc scale.

\section{Correlation Function of Dwarf Galaxies}
Besides luminosity, the bump should affect the large-scale structure of galaxies. \cite{Pilipenko24} showed that at high redshifts, the large-scale structure in the bump models differ significantly from the standard spectra. To test bump effects at low redshift, we measure the correlation function $\xi$ for model galaxies with $M_B < -14$ in both models, using the estimator:
\begin{equation}
\xi(r) = \frac{DD(r)}{RR(r)} - 1,
\label{eq:corfunc}
\end{equation}
where $DD$ is the number of galaxy pairs with separations in bin centered at $r$, and $RR$ is the number of pairs for randomly distributed points in the same bin.

Correlation functions were measured from 200 kpc to 7 Mpc. At very small separations, galaxy pairs cannot exist due to merging. GRUMPY treats halos closer than the virial radius of the larger halo as one halo. Due to the small simulation box size ($\sim50$ Mpc), the correlation function is distorted for scales larger than $L_{box}/4$, so we limit analysis to smaller scales. The bump scale corresponds to a linear size of 1.3 Mpc, so scales near 1 Mpc are most interesting.

\begin{figure}
    \centering
    \includegraphics[width=0.5\textwidth]{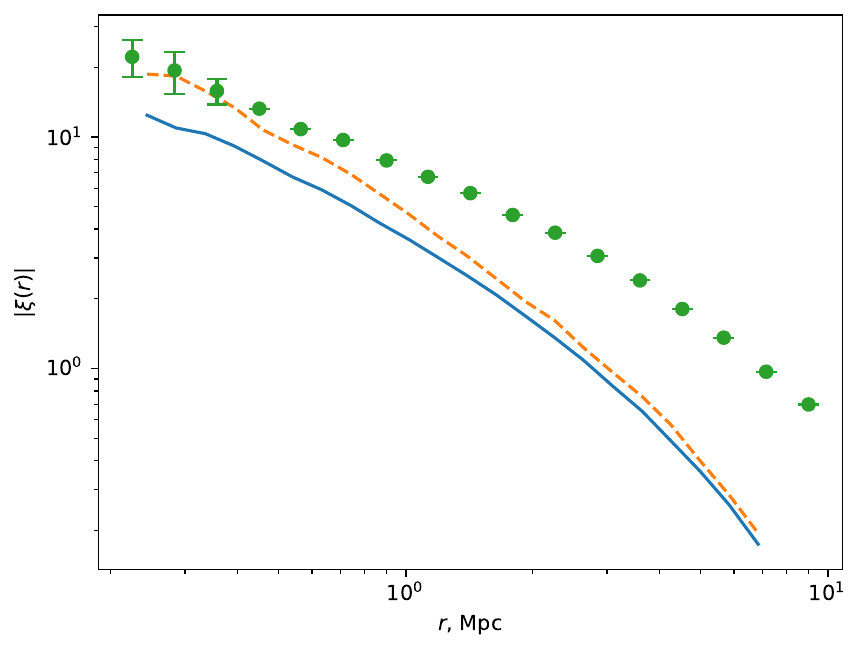}
    \caption{Correlation function of galaxies with $M_B < -14$ in bump (solid line) and no-bump (dashed line) models. Points show measured correlation function from 2dF catalog for galaxies with $M_B < -20$.}
    \label{fig:corfunc}
\end{figure}

Figure~\ref{fig:corfunc} shows that correlation functions for bump and no-bump models are similar. Correlation function in the bump model doesn't have any features and is about 1.5 times lower at small scales 0.2–0.3 Mpc. For comparison, points show the correlation function for galaxies with $M_B < -20$ from the 2dF catalog \cite{Hawkins03}. The observed function differs significantly, likely because 2dF galaxies are much brighter and probe larger-scale structures than our models.

To our knowledge, despite catalogs of dwarf galaxies do exist (e.g., \cite{Karachentsev13}), observed dwarf galaxy correlation functions have not yet been obtained. It is due to difficulties in constructing random catalogs for computing RR value from \ref{eq:corfunc}, because such catalog must satisfy all galaxy selection criteria, which are used in observational catalog construction.

\section{Observational Constraints on Bump Parameters}
The strong distortion of the dwarf galaxy luminosity function at $-17 < M_B < -13$ in the bump model disagrees with observed luminosity functions \cite{klypin15}. Under GRUMPY assumptions, the bump significantly affects stellar masses of present-day galaxies, and a bump at $k_0=4.7$ Mpc$^{-1}$ with amplitude $\mathcal{A}=2$ is ruled out by nearby galaxy observations.

Using observational data, we constrain allowed bump parameters. Numerical modeling on a grid of $k_0$, $\mathcal{A}$ is resource-intensive, so we use results from \cite{Yankelevich18}, which show small cosmological model variations produce linear changes in halo mass functions. We interpolate our results in parameter space between $\mathcal{A}=2$ and no bump $\mathcal{A}=0$, assuming the luminosity function depends on bump amplitude as:
\begin{eqnarray}
    \Phi(M_B;\mathcal{A}) = \Phi(M_B;0) + \nonumber \\ 
 + \mathcal{A} \frac{\Phi(M_B;2)-\Phi(M_B;0)}{2}
 \label{eq:phi-a}
\end{eqnarray}

Using observed luminosity functions and errors from \cite{klypin15} for $M_B < -14$ (the complete sample), we compute $\chi^2$ and find the probability. With one free parameter $\mathcal{A}$, the probability exceeds 0.003 (corresponding to 3$\sigma$) for $\mathcal{A} < 0.25$, which sets an upper limit on the bump amplitude.

\begin{figure}
    \centering
    \includegraphics[width=\linewidth]{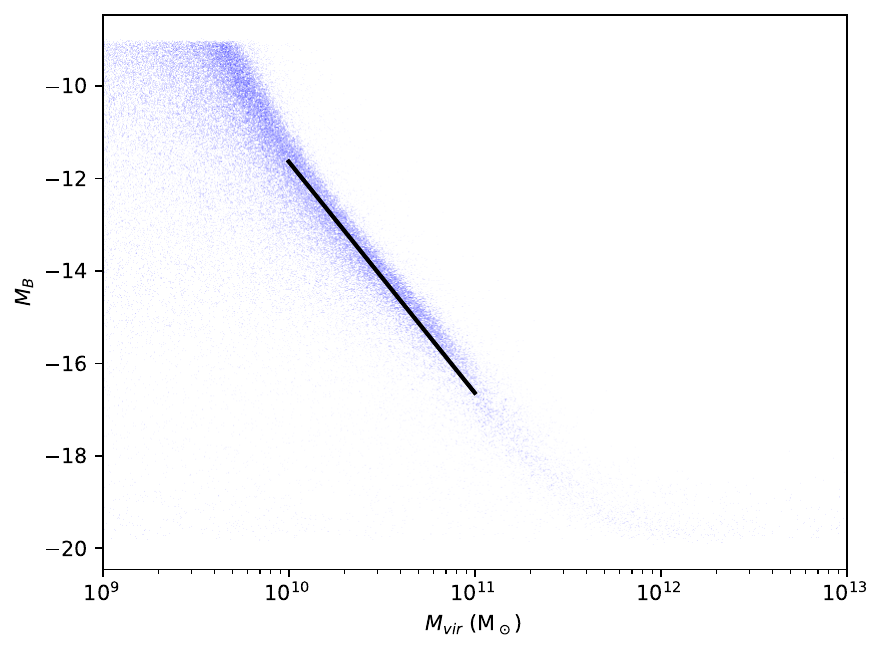}
    \caption{Mass-luminosity relation for galaxies in both models. The line shows $L \propto M_v^2$.}
    \label{fig:ML}
\end{figure}

To assess $k_0$ influence on the luminosity function, we analyze the mass-luminosity relation shown in Fig.~\ref{fig:ML}. It is identical for both models, i.e., independent of the cosmological power spectrum. In the luminosity range $-17 < M_B < -13$ where the bump appears, luminosity scales as the square of the virial mass. The bump will not contradict observations if this feature shifts to $M_B > -14$, a poorly constrained region observationally. This requires luminosities of galaxies with $M_B = -17$ formed by the bump to decrease by a factor of 16, implying masses decrease by a factor of 4 (from Fig.~\ref{fig:ML}). According to (\ref{eq:m-k0}), this corresponds to a bump wavenumber $k_0 = 7.4$ Mpc$^{-1}$ or wavelength 0.8 Mpc. Thus, current observations constrain the bump only for $k_0 < 7.4$ Mpc$^{-1}$.

The bump effect at smaller wavelengths may be more complex than this estimate, since dwarf galaxies with masses $\lesssim 10^{10} M_\odot$ will be part of larger galaxies with $M_B < -14$. Further simulations with different $k_0$ are needed.

\section{Conclusion}
We studied dwarf galaxy luminosity distributions in two models: one with the standard initial perturbation spectrum, the other with a bump at 1.3 Mpc wavelength and amplitude $\mathcal{A} = 2$ described by (\ref{eq:bump}). Model catalogs with luminosities were obtained by applying semi-analytic model GRUMPY to the results of cosmological numerical simulations.

%Using cosmological simulations to predict dark matter halo distributions and growth histories, we applied the semi-analytic GRUMPY model to compute galaxy luminosities.

The main result — luminosity functions in both models — is shown in Fig.~\ref{fig:lum}. The bump with amplitude $\mathcal{A} = 2$ is inconsistent with observations. Interpolating between bump and no-bump models yields an upper limit $\mathcal{A} < 0.25$ at 3$\sigma$ confidence. A slight decrease in bump wavelength from 1.3 to 0.8 Mpc shifts the luminosity function feature, caused by bump, into a poorly constrained observational region. The galaxy correlation function for $M_B < -14$ on scales below 1 Mpc depends on the initial spectrum but has not yet been measured observationally for such galaxies.

Future work includes improving the semi-analytic model to include supermassive black hole growth and star accretion during galaxy mergers, and simulating models with different bump parameters. Since a large fraction of the Local Volume is occupied by the Local Void, which may bias luminosity functions, we plan to study luminosity functions in overdense and underdense regions.

\section{Acknowledgments}
The work of O.R. Skorikov and S.V. Pilipenko was supported by the Basis Foundation grant for theoretical physics and mathematics development. The work of S.A. Drozdov was supported by grant NNG-36-2024.

\bibliographystyle{ieeetr}
\bibliography{refs}

\end{document}